\documentclass[]{spie}  

\usepackage{enumerate} 
\usepackage{amsmath,amsfonts,amssymb}
\usepackage{graphicx}
\usepackage[colorlinks=true, allcolors=blue]{hyperref}
\usepackage{multirow}
\usepackage{booktabs}
\usepackage{xcolor}
\usepackage{siunitx}
\title{Preliminary results of sky brightness measurements in near-infrared at Lenghu, China}

\author[a]{Jinji Li}
\author[a]{Bin Ma}
\author[a]{Zhongnan Dong}
\author[a]{Haoran Zhang}
\affil[a]{School of Physics and Astronomy, Sun Yat-sen University, Zhuhai, China}

\authorinfo{Further author information: (Send correspondence to B.M.) \\B. M.: E-mail: mabin3@mail.sysu.edu.cn}

\begin{document} 
\maketitle

\begin{abstract}
Low sky brightness is crucial for ground-based astronomical observations, because it limits the observational capability to detect fainter sources. Lenghu, located on the Tibetan Plateau in China, has been identified as an high-quality astronomical site in China, including dark sky in optical band. In this work, we will report the preliminary results of  near-infrared sky brightness measurements at Lenghu. Utilizing a wide-field small telescope equipped with an InGaAs camera, we have been conducting long-term monitoring of near-infrared sky brightness in the $J$ and $H’$ bands, respectively, since January 2024. For each  image, photometry and astrometry were performed, then sky background was calibrated by standard stars from the 2MASS catalog. This report includes preliminary results on the sky brightness at zenith in the $J$ and $H’$ bands, as well as their variations with solar elevation at Lenghu. Our initial results indicate that the near-infrared sky brightness at Lenghu is comparable to that of other world-class sites, and long-term monitoring will be continued.
\end{abstract}


\section{INTRODUCTION}
\label{sec:intro}  

There are a few key parameters essential for assessing the quality of an astronomical site. For ground-based near-infrared (NIR) and optical astronomical observations, preferred locations typically feature good seeing, high clear fraction, high atmospheric transparency and low sky brightness.\cite{krisciunas1997optical} Sky brightness predominantly determines the signal-to-noise ratio of faint sources. This is particularly evident in the NIR band, where the sky is  $\sim100$  times brighter than in the optical band.\cite{sivanandam2012characterizing} Consequently, conducting a NIR night sky brightness survey at a newly identified site is essential. Such a survey provides invaluable insights and guidance for the operation and planning of telescopes currently installed or proposed to be situated at the site.

The emission mechanisms of the sky background in the NIR band have been extensively studied. \cite{leinert19981997} Within this band, the major contributors to sky brightness are integrated starlight, zodiacal light, and OH airglow.  \cite{meinel1950oh,meinel1950oh2} For ground-based observatories, the intensity of OH airglow significantly overshadows other sources, thus predominating the overall sky brightness. \cite{ramsay1992non} Due to the characteristics of OH airglow, the temporal and spatial distribution of NIR sky brightness varies. 

Lenghu, located on the Tibetan Plateau, has been recognized as an high-quality astronomical observatory site in China, characterized by a median seeing of $0.75^{\prime \prime}$, a clear fraction of 70\%, and an optical sky brightness of 22 mag arcsec$^{-2}$. \cite{deng2021lenghu} Four telescopes, including the Wide Field Survey Telescope,  have already begun scientific operations there.\cite{chen2023basic} However, no measurements of the NIR night sky brightness had been previously conducted, which represents a significant limitation for the site's further development. To demonstrate the NIR observational capabilities, we have been conducting long-term monitoring of the sky brightness in the $J$ and $H’$  bands at Lenghu. This monitoring can serve not only as a site testing for Lenghu but also for the study of NIR sky brightness characteristics across wide-band imaging. 

In this paper, we characterize the preliminary results of the NIR sky brightness monitoring initiated from January at Lenghu. Section 2 presents the instruments used for measurements, the processing of observational images and the methods for calculating sky brightness. Section 3 describes the preliminary results from our analysis. In Section 4, the preliminary NIR sky brightness results at Lenghu are compared with other world-class sites, and a discussion has been provided. 

\section{OBSERVATIONS and METHODS
}
\begin{figure} [htp]

\begin{center}
\includegraphics[scale=0.3]{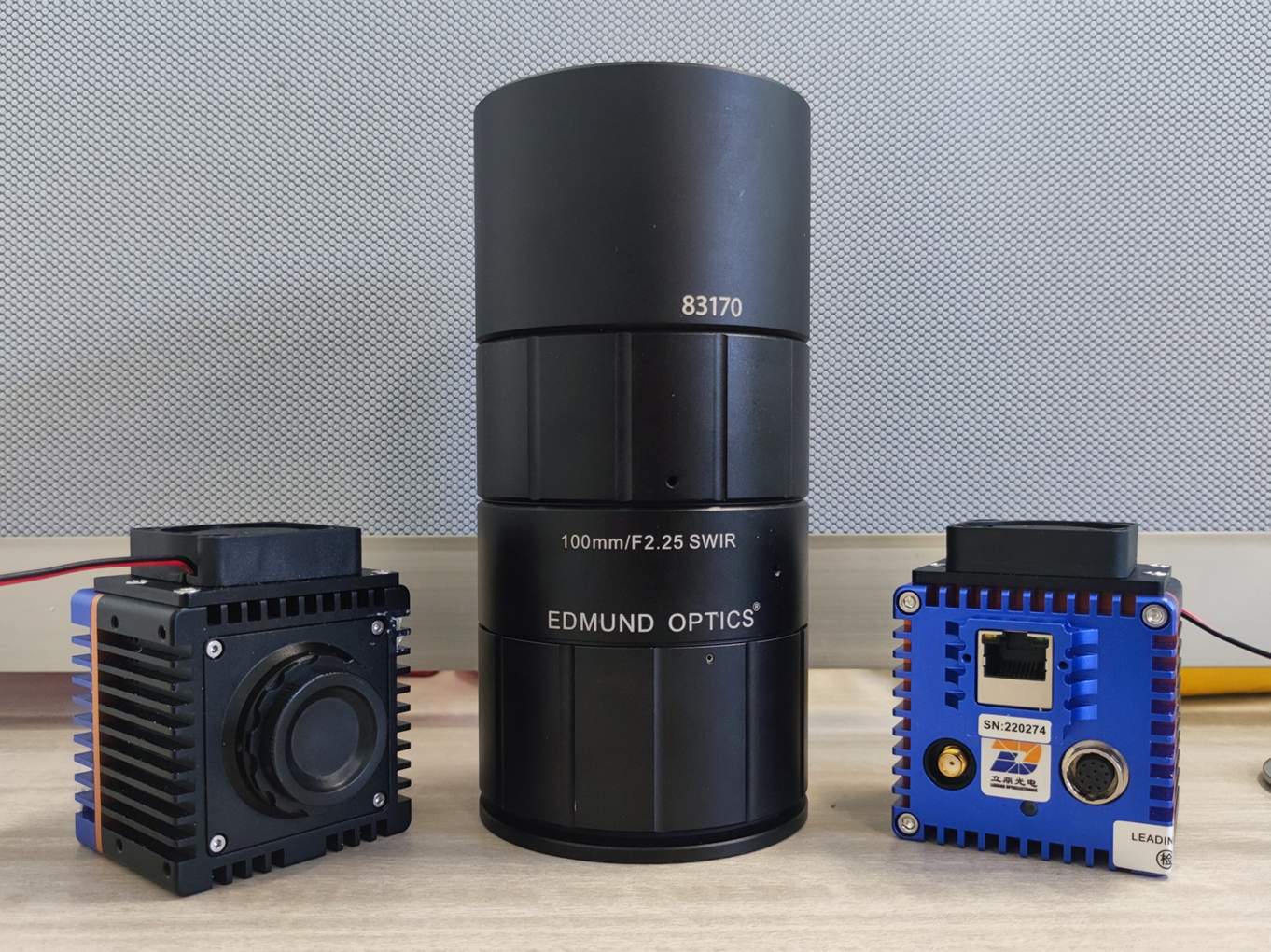}
\end{center}
\caption{The lens and camera used for sky brightness measurements} 
\label{fig:telescope}
\end{figure}

We employed a lens and an Indium Gallium Arsenide (InGaAs) camera at Lenghu to measure the zenith’s NIR sky brightness, as shown in Figure \ref{fig:telescope}. The observation equipment was installed at the 3850 m site on Saishiteng Mountain in Lenghu, which is the location for the MASTA project. \cite{gao2022nighttime,YuyiZHUANG}  We captured images of the sky background and stars, calibrating the sky brightness by the stars within the field of view.

\subsection{Observations}
We conducted $J$-band sky brightness monitoring from January to March, then from April, we switched to an $H’$-band filter, with monitoring still ongoing. The specifications of the equipment and observations are listed in Table \ref{tab:equipment}.  

\begin{table}[htp] 
\caption{Equipment and observations} \label{tab:equipment} \begin{center} \begin{tabular}{@{}llll@{}} \toprule & \textbf{Features} & \multicolumn{2}{c}{\textbf{Specifications}} \\ \midrule \multirow{2}{*}{Lens}& Aperture  & \multicolumn{2}{c}{4.4 cm} \\ & Focal Length  & \multicolumn{2}{c}{10 cm} \\ \hline \multirow{5}{*}{InGaAs camera}& Pixel number& \multicolumn{2}{c}{640 × 512} \\
 & Pixel size& \multicolumn{2}{c}{15 \si{\micro\meter} × 15 \si{\micro\meter}}\\ & Pixel Scale  & \multicolumn{2}{c}{31 arcsec pixel$^{-1}$} \\ & Field of View  & \multicolumn{2}{c}{5.51 × 4.41 deg$^{2}$} \\
 & Response range& \multicolumn{2}{c}{0.9 \si{\micro\meter} -- 1.7 \si{\micro\meter}}\\ \hline \multirow{2}{*}{Filter}&  & \multicolumn{2}{c}{$J$-band (1.09 \si{\micro\meter} -- 1.35 \si{\micro\meter})} \\
 & & \multicolumn{2}{c}{$H'$-band (1.48 \si{\micro\meter} -- 1.65 \si{\micro\meter})}\\
\hline  
 & Observation time& \multicolumn{2}{c}{$J$-band (2024.01 -- 2024.03)}\\
 & & \multicolumn{2}{c}{$H'$-band (2024.04 -- ongoing)}\\
 \hline 
 & Exposure time& \multicolumn{2}{c}{3 s}\\
 \hline 
\end{tabular}
 \end{center} 
\end{table}

\begin{figure} [htp]

\begin{center}
\includegraphics[scale=0.6]{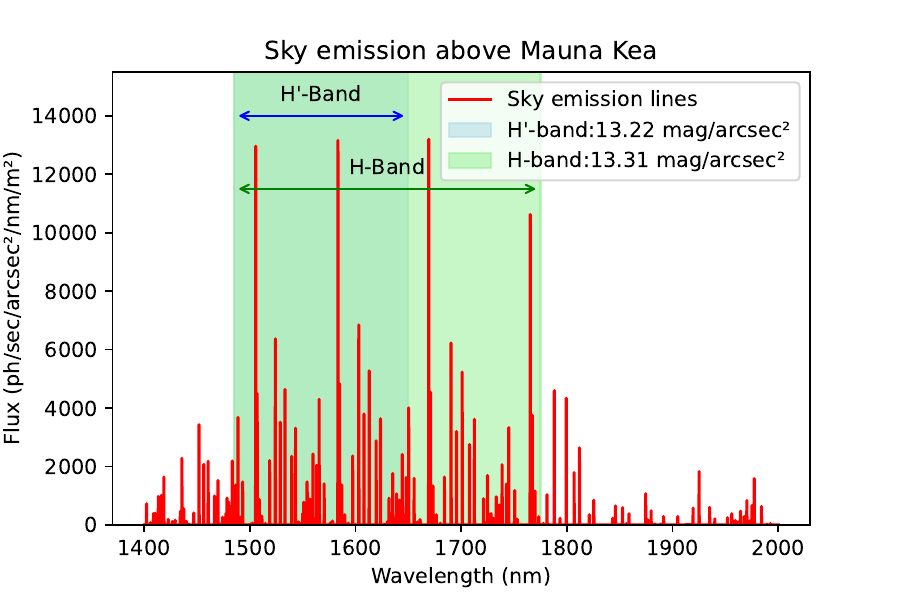}
\end{center}
\caption{The difference in sky brightness between the $H$-band and $H'$-band. Sky emission lines data above Mauna Kea is obtained from Gemini Observatory (\protect\url{https://www.gemini.edu/sciops/ObsProcess/obsConstraints/atm-models/nearIR_skybg_16_15.dat}).}

\label{fig:H'band}
\end{figure}

To measure the NIR sky brightness, we utilized an InGaAs camera, which has a spectral response range of 0.9 -- 1.7 microns, covering the entire standard $J$-band and part of the standard $H$-band (we call it $H'$-band hereafter) . To compare the differences in sky brightness between the $H$-band and $H'$-band, we analyzed the sky emission obtained at Mauna Kea and calculated the sky brightness for both the  $H$-band and $H'$-band (Figure \ref{fig:H'band}). \cite{lord1992new}  The sky brightness in the $H'$-band (13.22 mag arcsec$^{-2}$) is slightly brighter than in the $H$-band (13.31 mag arcsec$^{-2}$), but the difference is minimal. Therefore, the $H'$-band sky brightness measurements taken at Lenghu can be directly compared with $H$-band from other sites.

We fixed the telescope's pointing to the zenith to measure the NIR sky brightness, making the selection of an appropriate exposure time crucial. We ultimately set the exposure time to 3 seconds for two main reasons. First, to avoid star trailing and thereby prevent errors during the flux calibration. Second, due to the relatively high dark current and the NIR sky brightness, longer exposure times can easily lend to saturation. With a 3 seconds exposure time, we can measure the NIR night sky brightness when the solar elevation is lower than -7°. 

\subsection{Image reduction}
\begin{figure} [htp]

\begin{center}
\includegraphics[scale=0.6]{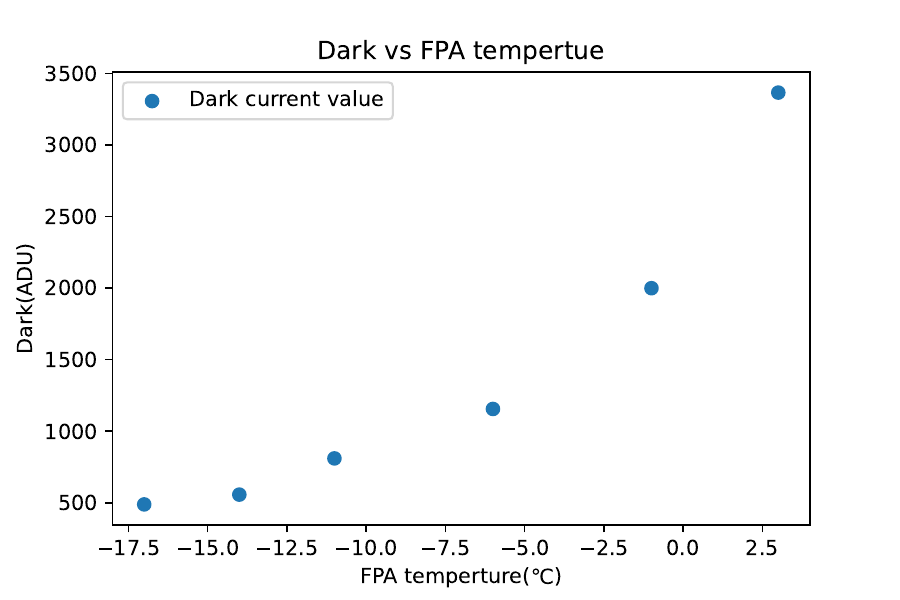}
\end{center}
\caption{Dark current values of InGaAs camera at different FPA temperatures. } 
\label{fig:dark}
\end{figure}

After obtaining the observation images, we first need to preprocess them: subtracting the bias and dark current, and correcting flat-fielding. During nighttime, the Focal Plane Array (FPA) temperature exhibits slight fluctuations due to changes in ambient temperature, which causes variations in the dark current of the InGaAs camera (Figure \ref{fig:dark}). For example, the dark current value demonstrates a change of 250 Analog-to-Digital Unit (ADU) as the FPA temperature shifts from -14°C to -11°C. Compared to the sky background level (600 ADU) , this variation is non-negligible. Due to fluctuations in the FPA temperature, to ensure consistency in temperature between the observed image and the subtracted dark frame, we need to scale the dark frame to prevent inaccurate measurements of sky brightness. 

\begin{figure} [htp]

\begin{center}
\includegraphics[scale=0.54]{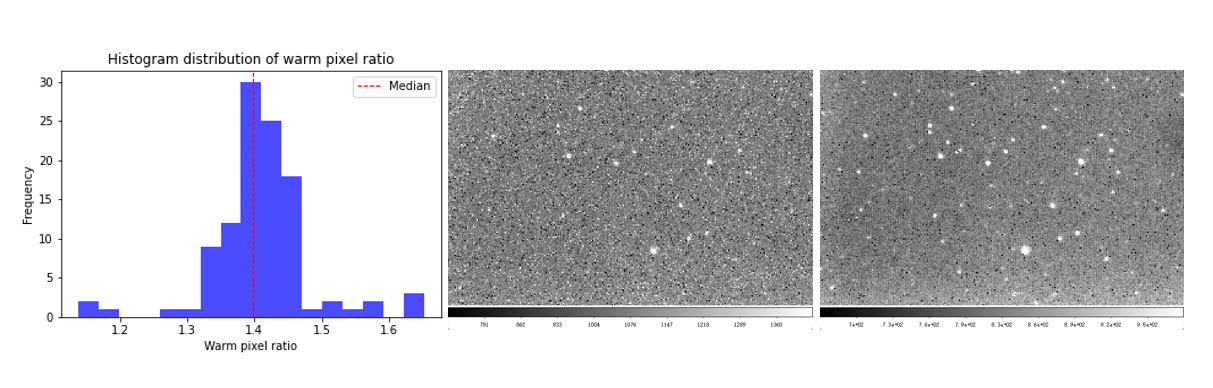}
\end{center}
\caption{Left: Distribution of the ratio of the differences between warm pixels and normal pixels. We use the median ratio to scale the dark current. Middle: Image after subtracting the unscaled dark current. Right: Image after subtracting the scaled dark current.} 
\label{fig:warm pixel}
\end{figure}

To achieve a more accurate subtraction of dark current, we scaled the dark frame to match science image. Within a specific temperature range, the magnitude of dark current pixel values demonstrates a linear relationship with temperature variations. Consequently, the scaling factor can be determined using the ratio of dark current values in the observed image to those in the dark frame. However, in the observed image, pixel values contain both dark current and sky background, which makes a direct comparison with the dark frame infeasible. Nevertheless, since the contribution of sky background is consistent across adjacent pixels in the observed image, the difference between warm and normal pixels does not include the contribution of sky background. Given that the scaling factors for warm and normal pixels are identical, we instead used the ratio of the differences between warm pixels and normal pixels  in the observed image and the dark frame for scaling the dark frame.  This scaling was then applied to the dark current images, and the results were remarkable (Figure \ref{fig:warm pixel}).

\subsection{Calibrations}
We employed Source Extractor for photometry and Scamp for astrometry.\cite{bertin1996sextractor,bertin2006automatic} Within Source Extractor, the FLUX\_AUTO parameter was chosen to obtain the flux of stars, utilizing the automatic aperture photometry routine derived from Kron's "first moment" algorithm.\cite{kron1980photometry} In the photometric results, we filtered stars with FLAG=0 and used the median of the sky background image as the typical sky background value. 

After performing star catalog cross-matching, we calibrated the zero-point of the image using the Two Micron All-Sky Survey (2MASS) star catalog as our reference.\cite{skrutskie2006two} The zero-point for each star was determined using the formula: 

\begin{equation}
\label{eq:zeropoint}
m_{\text{zero-point}} = 2.5 \times \log_{10}(\text{flux}) + m_{\text{2MASS}},
\end{equation}
in this equation, \( m_{zero-point} \) represents the zero-point magnitude, \( flux \) stands for the flux of stars, and \( m_{2MASS} \) refers to the magnitude with reference to 2MASS star catalog. 

\begin{figure} [htp]

\begin{center}
\includegraphics[scale=0.6]{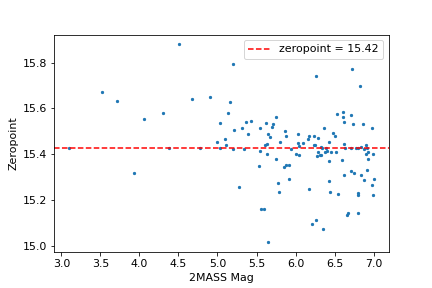}
\end{center}
\caption{Image zero-point calibration} 
\label{fig:zeropoint}
\end{figure}

We subsequently determined the zero-point of the image by taking the median of all matched stars, as depicted in Figure  \ref{fig:zeropoint}. 

Following the establishment of the image zero-point, we computed the sky brightness using the equation: 

\begin{equation}
\label{eq:ZEROPOINT}
m_{\text{sky}} = m_{\text{zero-point}} -2.5 \times \log_{10}(\text{B}) +2.5 \times \log_{10}(\text{scale$^{2}$}) ,
\end{equation}

where \( m_{sky} \) represents the sky brightness (in mag arcsec$^{-2}$), \( m_{zero-point} \) is the image zero-point,  \( B\) is the sky background value (in ADU), and  \( scale\) is the pixel scale (in arcsec pixel$^{-1}$ ).

In our measurement system, with a limiting magnitude of approximately 8th magnitudes, we calculated the integrated starlight for stars beyond this limiting magnitude ($\sim20$ mag arcsec$^{-2}$). Compared to the sky brightness ($\sim16$ mag arcsec$^{-2}$) , the flux contribution from stars fainter than the limiting magnitude is negligible. 

\section{RESULTS
}
\begin{figure} [h]

\begin{center}
\includegraphics[scale=0.4]{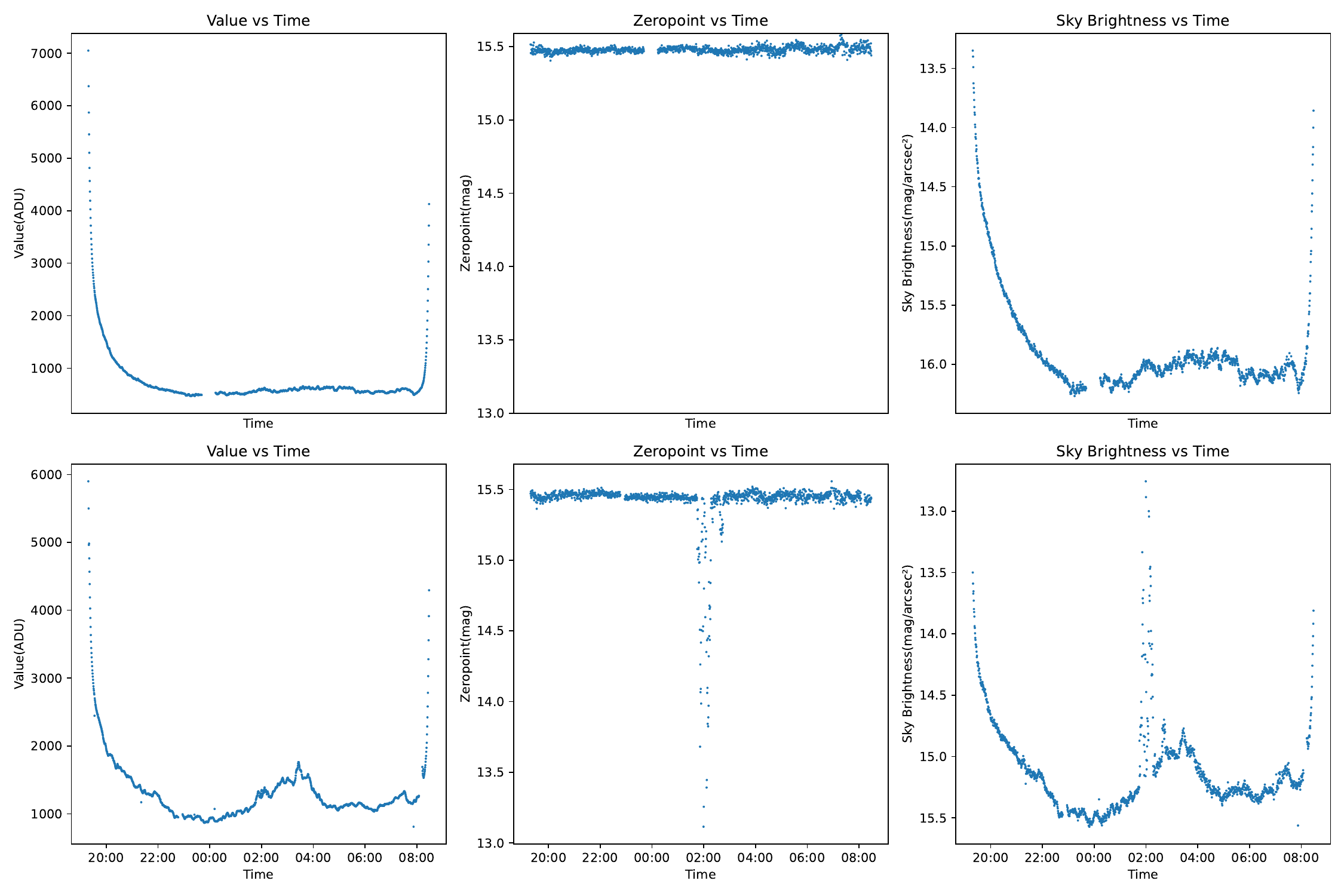}
\end{center}
\caption{Variations of sky background values (left, in ADU), image zero-point magnitude (middle), and sky brightness (right, in mag arcsec$^{-2}$). Upper: On a clear, moonless night. Lower: A decrease in image zero-point when the measured sky region is obscured by clouds. } 
\label{fig:sky brightness}
\end{figure}

In Lenghu, from January to March, we conducted measurements of the sky brightness in the $J$-band. Figure \ref{fig:sky brightness} illustrates two typical observational scenarios. On a moonless night (upper panels in Figure \ref{fig:sky brightness}) , the zero-point magnitude value remains relatively constant, indicating that it is a completely clear night. On a such night, the sky brightness gradually decreases after sunset until it reaches a relatively stable level, which is then maintained until it begins to rise again at sunrise. Generally, we use the sky brightness of the relatively stable phase to represent the level of sky brightness for that night. When the measurement region is obscured by clouds (bottom panels in Figure \ref{fig:sky brightness}) , it results in  a decrease in the zero-point values. 

\begin{figure} [h]

\begin{center}
\includegraphics[scale=0.6]{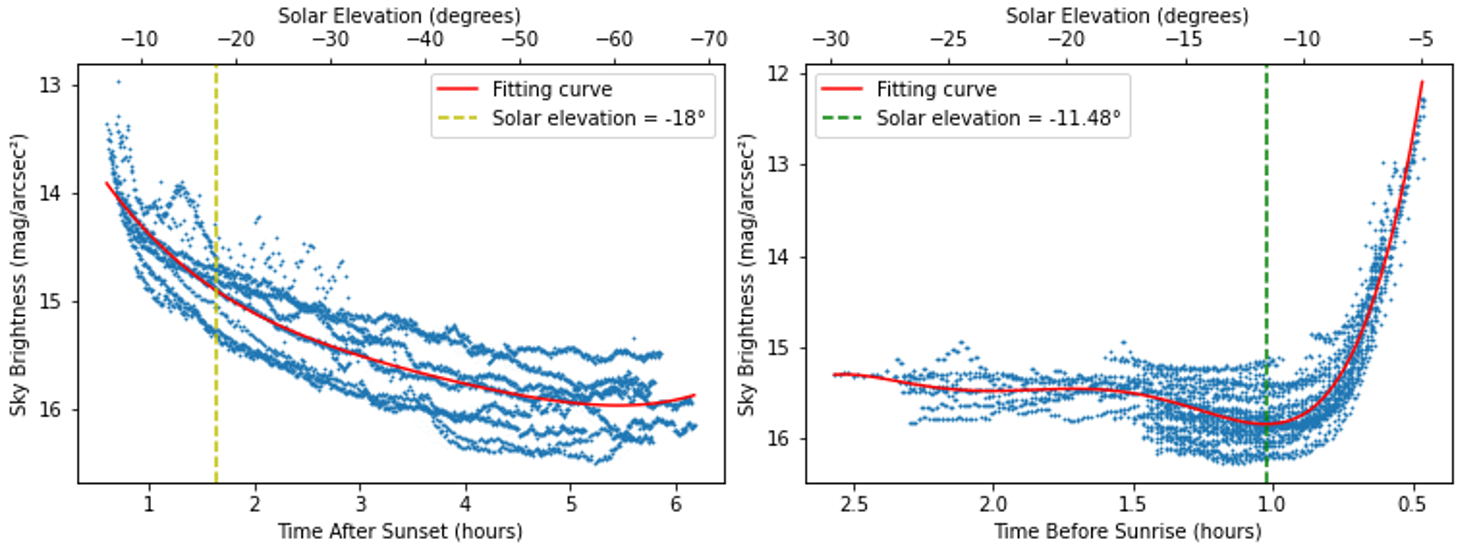}
\end{center}
\caption{Variations of sky brightness after sunset and before sunrise. Left: Sky brightness continue to darken until $\sim5$ hours after sunset. Right: Sky starts to brighten $\sim1$ hour before sunrise, following a brief dimming.} 

\label{fig:set and rise}
\end{figure}

We analyzed the variations in sky  brightness after sunset and before sunrise, respectively, by synthesizing multiple nights of observational data (Figure \ref{fig:set and rise}) . After sunset, the sky continues to darken for about five hours until it reaches a relative stability state. This is different from that in the optical band, where the sky brightness achieves relative stability more quickly, once the solar elevation descends below -18°. Additionally, while changes in sky brightness are symmetric around sunset and sunrise in the optical band, they exhibit asymmetry in the $J$-band. Specifically, sky brightness in the $J$-band starts to increase about one hour before sunrise, following a slight decline. 

\begin{figure} [h]

\begin{center}
\includegraphics[scale=0.4]{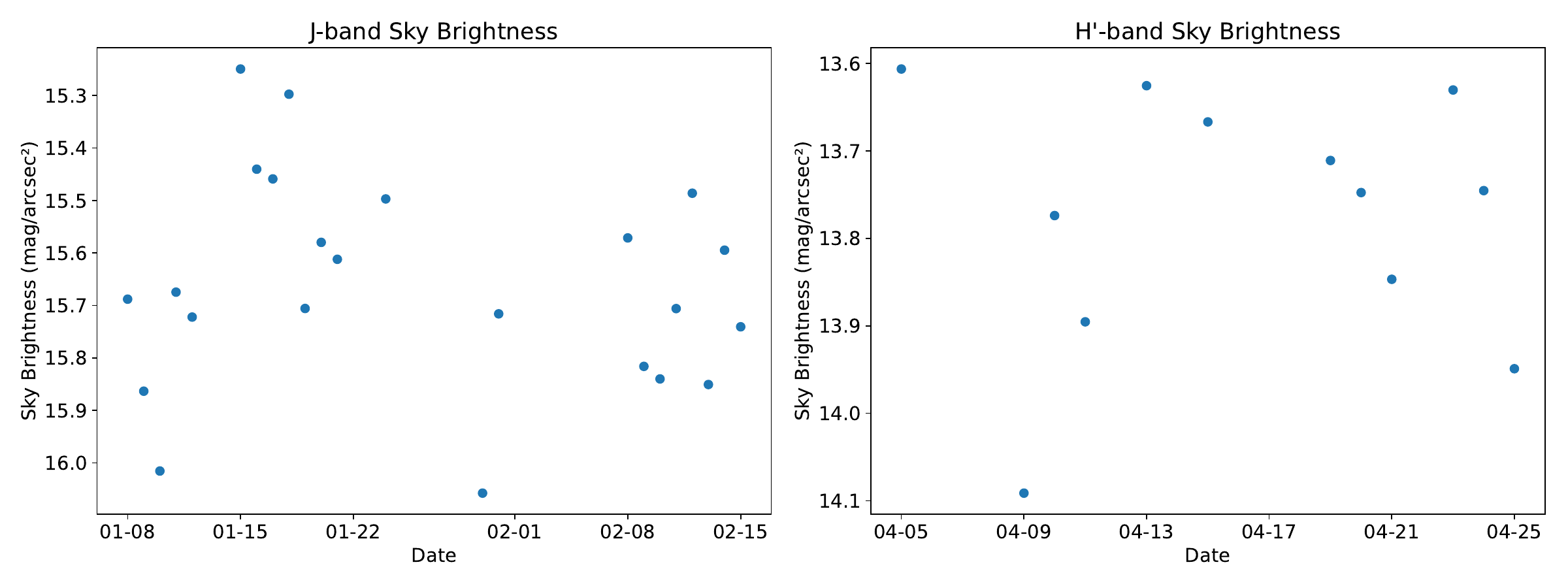}
\end{center}
\caption{Night sky brightness during monitoring period. Left: $J$-band. Right: $H'$-band.}
\label{fig:monitoring}
\end{figure}

We use the median value of the relatively stable phase at night to represent the sky brightness of that evening. Figure \ref{fig:monitoring}  shows the sky brightness in the $J$ and $H'$ bands, respectively, for each night during the monitoring period at Lenghu. For the $J$-band, there were 23 days of observations, with sky brightness ranging from 15.2 to 16.1 mag arcsec$^{-2}$. For the $H'$-band, there were 12 days of observations, with sky brightness ranging from 13.6 to 14.1 mag arcsec$^{-2}$. 

\section{SUMMARY and DISCUSSIONS
}
We measured the near-infrared sky brightness in the $J$ and $H'$ band at Lenghu via imaging. During the preprocessing of the images, we paid particular attention to the accuracy of dark current subtraction and employed warm pixel ratio scaling for dark frame. Based on our preliminary results, the levels of sky brightness in Lenghu are comparable to those at other world-class astronomical sites (Table  \ref{tab:observatory}).

\begin{table}[h] 
\caption{The near-infrared sky brightness of  world-class sites} \label{tab:observatory} \begin{center} \begin{tabular}{@{}lllll@{}c} \toprule Site& Altitude (m)& Type& $J$ ( mag arcsec$^{-2}$)& $H$ ( mag arcsec$^{-2}$)&Reference\\ \midrule 
 Lenghu& 3800& Preliminary results& 15.2 -- 16.1& 13.6 -- 14.1&This work\\\multirow{2}{*}{Mauna Kea}& 4200& Darkest& 16.75& 14.75& (1)\\ & & Average& 15.6& 14.0& \\  La Palma& 2500& Average& 15.5& 14.0& (1)\\ Paranal& 2635&  Darkest& 16.5& 14.4& (1)(2)\\
 South Pole& 2835& --& 16.8 -- 16.0& 15.2 -- 14.2&(3)\\ Cerro Pachon& 2200& --
& 16.0& 13.9& (1)\\
 Mt.Hamilton& 1283& --& 16.0& 14.0&(1)\\
 Kitt Peak& 2096& --& 15.7& 13.9&(1)\\
 Anglo Australian Obs.& 1164& --& 15.7& 14.1&(1)\\ \hline 
 & & & & &\\ \end{tabular}
 Note: (1) S{\'a}nchez et al. (2008) \cite{sanchez2008night} and references therein, (2) Cuby et al. (2000) \cite{cuby2000isaac} , (3) Phillips et al. (1999)\cite{phillips1999near}
 \end{center} 
\end{table}

In the $J$ and $H'$ bands, the sky brightness is primarily sourced from the OH airglow\cite{sivanandam2012characterizing}. This emission is highly variable on short (minutes) and longer (hours to years) timescales. On a short timescale, variations have been confirmed to be caused by changes in the densities of reactants that produce the OH molecule, due to the propagation of gravity waves in the high atmosphere.\cite{ramsay1992non,swenson1994oh} Diurnal variations over longer timescales, along with correlations with the solar cycle, have been observed. Studies have shown that the OH airglow is brighter during periods of intense solar activity\cite{pertsev2008response}. Our monitoring of sky brightness occurs near the solar maximum, which probably leads to an increase in the brightness of the near-infrared sky. In the future, based on our long-term monitoring results, we will analyze the temporal and spatial characteristics of near-infrared wide-band sky brightness.

In June 2024, an 80cm near-infrared telescope will be installed at Lenghu, operated in the $J$ and $K$ bands. Therefore, our monitoring of the $K$-band sky brightness will be conducted from June.

\acknowledgments 
 B.M.  acknowledges support from the National Natural Science Foundation of China under grant No. 11733007.

\bibliography{report} 
\bibliographystyle{spiebib} 

\end{document}